\documentclass[12pt]{article}

\input{tcilatex}

\begin{document}

\begin{center}
\textbf{Completely Integrable Models of Non-linear Optics}

\vspace{0.5in} Andrey I. Maimistov

\vspace{0.2cm} Department of Solid State Physics, Moscow Engineering Physics
Institute,

Moscow, 115409, Russia

electronic address: maimistov@pico.mephi.ru

\bigskip

\textbf{ABSTRACT}
\end{center}

\bigskip

The models of the non-linear optics in which solitons were appeared are
considered. These models are of paramount importance in studies of
non-linear wave phenomena. The classical examples of phenomena of this kind
are the self-focusing, self-induced transparency, and parametric interaction
of three waves. At the present time there are a number of the theories based
on completely integrable systems of equations, which are both generations of
the original known models and new ones. The modified Korteweg-de Vries
equation, the non- linear Schrodinger equation, the derivative non-linear
Schrodinger equation, Sine-Gordon equation, the reduced Maxwell-Bloch
equation, Hirota equation, the principal chiral field equations, and the
equations of massive Thirring model are gradually putting together a list of
soliton equations, which are usually to be found in non-linear optics theory.

\bigskip

Keywords: solitons, self-induced transparency, non-linear fibres, wave
interaction PACS number: 42.65.

\newpage

\section{Introduction}

A non-linear wave is one of the fundamental objects of nature. They are
inherent to aerodynamics and hydrodynamics, solid state physics and plasma
physics, optics and field theory, chemistry reaction kinetics and population
dynamics, nuclear physics and gravity. All non-linear waves can be divided
into two parts: dispersive waves and dissipative ones. The history of
investigation of these waves has been lasting about two centuries. In 1834
J.S. Russell discovered the extraordinary type of waves without the
dispersive broadening. In 1965 N.J. Zabusky and M. D. Kruskal found that the
Korteweg-de Vries equation has solutions of the solitary wave. This solitary
wave demonstrates the particle-like properties, i.e., stability under
propagation and the elastic interaction under collision to one another. For
these reasons these solitary waves were named solitons. In succeeding years
there has been a great deal of progress in understanding of soliton nature.
Now solitons have become the primary components in many important problems
of non-linear wave dynamics. It should be noted that non-linear optics is
the field, where all soliton features are exhibited to a great extent.

The self-focusing of light beams and self-induced transparency phenomenon
set the bright example of an important role of solitons in non-linear optics.

The non-linear wave propagation in non-resonant medium also occupies an
important place in non-linear optics. As c consequence of modulation
instability, continuous wave transforms to number of single pulses. These
pulses can under certain conditions evaluate into solitons. Nowadays the
optical solitons propagation in fibres has attracted considerable attention.

The other classical problem of non-linear optics is the parametric
interaction of waves. The harmonic generation, stimulated by both Raman and
Brillouin scattering, parametric amplifications, sum-frequency mixing, and
four-wave mixing have been the subject of many investigations in this field.
The three-wave interaction process is of particular interest because it
gives us a new instance of the application of the soliton occurrence in
non-linear system without dispersion.

There are cases of the Hamiltonian systems when the canonical transformation
makes the equations of motion trivially integrable after the conversion to
new variables. Then they say that Hamiltonian system admits the action-angle
variables. If the action-angle variables exist, then the Hamiltonian system
is \emph{completely integrable}. There is powerful tool to investigate the
integrable systems. It is Inverse Scattering Transform (IST) introduced by
Gardner, Greene, Kruskal and Miura (1967). The implementation of IST in
non-linear optics results in new examples of the optical soliton phenomena.
Moreover, this method leads to new analytical approaches for near integrable
systems.

\section{Classical examples}

Let us consider three phenomena mentioned above for that theories have been
based on completely integrable equations.

\subsection{Self-induced transparency}

A self-induced transparency (SIT) phenomenon consists in the propagation of
a powerful ultra-short pulse (USP) of light through a resonance medium
without the distortion and energy loss of this pulse \cite{R1, R2, R3, R4}.
This phenomenon is characterised by the continuous absorption and
re-emission of electromagnetic radiation by resonant atoms of medium in such
a manner that steady-state optical pulse propagates. In the ideal case the
energy dissipation of the USP is invisible and the state of the resonant
medium is not varying. In this means the medium is transparent. The group
velocity of such steady-state pulse, called $2\pi $-pulse or soliton of SIT,
is less than the phase speed of light in a medium. The group velocity
depends on a $2\pi $-pulse duration: the shorter is the duration, the higher
is its speed \cite{R2, R3, R4, R5}. When two pulses of the different
velocities spread in the medium, the second pulse may overtake the first and
a collision will take place. After the collision, the solitons keep their
shape and velocity (but in general all other parameters of solitons may
alter). This fundamental property of the SIT solitons has been studied many
times both theoretically and experimentally \cite{R3, R6, R7}.

From the mathematical point of view this property is a consequence of the
complete integrability of the reduced Maxwell-Bloch equations, describing
the SIT in the two-level media with non-degenerated levels \cite{R8} - \cite
{R13}. The $2\pi $-pulses answer the single- soliton solutions of these
equations, and the process of \ ''collision'' reflects the evolution of the
double-soliton solution -- its asymptotically transformation into a pair of
solitons under certain conditions (see, for example, \cite{R6, R9}, and \cite
{R14} - \cite{R15}).

The simplest theory describing the self-induced transparency phenomenon was
developed by McCall and Hahn \cite{R1, R2}. In general, the theory of the
interaction of radiation with an ensemble of two-level atoms is based on the
Bloch equations for atoms and the Maxwell equations for the classical
electromagnetic field. In an isotropic dielectric the set of Maxwell
equations reduced to one equation for the electric field $\vec{E}=\vec{l}E$.
For a plane wave with constant polarisation vector $\vec{l}$ one can obtain
the following system of total Maxwell-Bloch (MB) equations

\begin{equation}
\frac{\partial ^{2}E}{\partial z^{2}}-\frac{1}{c^{2}}\frac{\partial ^{2}E}{%
\partial t^{2}}=\frac{4\pi n_{A}d}{c^{2}}\left\langle \frac{\partial
^{2}r_{1}}{\partial t^{2}}\right\rangle ,  \label{eq1-1}
\end{equation}

\begin{equation}
\frac{\partial r_{1}}{\partial t}=-\omega _{a}r_{2},\quad \frac{\partial
r_{2}}{\partial t}=\omega _{a}r_{1}+\frac{2d}{\hbar }Er_{3},\quad \frac{%
\partial r_{3}}{\partial t}=-\frac{2d}{\hbar }Er_{2},  \label{eq1-2}
\end{equation}
where $d$ is the projection of a matrix element of the dipole operator on
the direction of $\vec{l}$, $n_{A}$ \ is the concentration of resonant
atoms. It should be noted that the components of Bloch vector $r_{1}$, $%
r_{2} $, and $r_{3}$\ depend on the atomic resonance frequency $\omega _{a}$%
. Hereafter the angular brackets represent summation over all the atoms
characterised by the frequency $\omega _{a}$.

The Bloch equations contain products of the field $E$ and the polarisations $%
r_{2}$\ and $r_{3}$ responsible for interference between the opposite
propagated waves. It has been shown \cite{R16, R17}, however, that if the
density of resonant atoms is small enough to make the parameter $4\pi
n_{A}d^{2}/\hbar \omega _{a}$ less than unity, interference may be
neglected. It was found that for a typical value of $d$ $\sim $ 1 Debye, $%
\omega _{a}\sim $~$10^{15}$ $s^{-1}$ and \ $n_{A}\ll 10^{23}cm^{-3}$ one may
not take into account the backward wave generation by a forward running
pulse. Thus, the MB equations convert into the simple \emph{reduced
Maxwell-Bloch} (RMB) system of equations:

\begin{equation}
\frac{\partial E}{\partial z}+\frac{1}{c}\frac{\partial E}{\partial t}=-%
\frac{2\pi n_{A}d}{c}\left\langle \frac{\partial r_{1}}{\partial t}%
\right\rangle ,  \label{eq2-1}
\end{equation}

\begin{equation}
\frac{\partial r_{1}}{\partial t}=-\omega _{a}r_{2},\quad \frac{\partial
r_{2}}{\partial t}=\omega _{a}r_{1}+\frac{2d}{\hbar }Er_{3},\quad \frac{%
\partial r_{3}}{\partial t}=-\frac{2d}{\hbar }Er_{2}.  \label{eq2-2}
\end{equation}
It should be emphasised that in both the MB and RMB equations symbol $E$
denotes the real value of the electric field strength. However, every so
often the electromagnetic wave can be represented as a quasi-monochromatic
one

\begin{equation}
E(z,t)=2A(z,t)\cos [k_{0}z-\omega _{0}t+\varphi (z,t)]=\mathcal{E}(z,t)\exp
[i(k_{0}z-\omega _{0}t)],  \label{eq3}
\end{equation}
where $\omega _{0}$ is the radiation frequency, $k_{0}$ is the wave number,
and the real envelope $A(z,t)$ and the phase $\varphi (z,t)$ are slowly
varying functions of $z$ and $t.$ This is an approximation which means that
the envelope and the phase obey the inequalities

\begin{equation}
|\partial A/\partial t|\ll \omega _{0}|A|,|\partial A/\partial z|\ll
k_{0}|A|,|\partial \varphi /\partial t|\ll \omega _{0}|\varphi |,|\partial
\varphi /\partial z|\ll k_{0}|\varphi |.  \label{eq4}
\end{equation}
Besides, the envelope amplitudes are usually so weak that Rabi frequency ($%
\max |dA/\hbar |$ ) turns out to be much less than the resonance transition
frequency. The resulting system of equations can be represented as

\begin{equation}
\frac{\partial q}{\partial z}+\frac{1}{c}\frac{\partial q}{\partial t}%
=-\alpha ^{\prime }\left\langle P\right\rangle ,\quad q\left( \frac{\partial
\varphi }{\partial z}+\frac{1}{c}\frac{\partial \varphi }{\partial t}\right)
=\alpha ^{\prime }\left\langle Q\right\rangle ,  \label{eq5-1}
\end{equation}

\begin{equation}
\frac{\partial Q}{\partial t}=\left( \Delta \omega +\frac{\partial \varphi }{%
\partial t}\right) P,\quad \frac{\partial P}{\partial t}=-\left( \Delta
\omega +\frac{\partial \varphi }{\partial t}\right) Q+qR,\quad \frac{%
\partial R}{\partial t}=-qP.  \label{eq5-2}
\end{equation}
where $\Delta \omega =(\omega _{a}-\omega _{0})$, $\alpha ^{\prime }=2\pi
\omega _{0}n_{A}d^{2}/\hbar c$, $q=dA/\hbar $\ is a normalised slowly
varying pulse envelope and $P$, $Q$ and $R$ are connected with initial Bloch
vector components by the relations

\begin{eqnarray*}
r_{1} &=&P(z,t)\sin [k_{0}z-\omega _{0}t+\varphi (z,t)]+Q(z,t)\cos
[k_{0}z-\omega _{0}t+\varphi (z,t)],\quad \\
r_{3} &=&-R(z,t)
\end{eqnarray*}

To describe the SIT phenomenon McCall and Hahn used the system (\ref{eq5-1})
and (\ref{eq5-2}). Furthermore, if we confine the analysis to situations
when the input optical pulse does not carry any phase modulation, i.e. $%
\partial \varphi /\partial z=\partial \varphi /\partial t=0$\ at z = 0 and
the form factor of the inhomogeneous line is a symmetrical function of
frequency detuning $\Delta \omega $, then equations (\ref{eq5-1}) and (\ref
{eq5-2}) yield $\partial \varphi /\partial z=\partial \varphi /\partial t=0$%
\ at any $z$ and $t$. In this case equations (\ref{eq5-1}) and (\ref{eq5-2})
reduce to the system of \emph{SIT equations}

\begin{equation}
\frac{\partial q}{\partial z}+\frac{1}{c}\frac{\partial q}{\partial t}%
=-\alpha ^{\prime }\left\langle P\right\rangle ,  \label{eq6-1}
\end{equation}

\begin{equation}
\frac{\partial Q}{\partial t}=\Delta \omega P,\quad \frac{\partial P}{%
\partial t}=-\Delta \omega Q+qR,\quad \frac{\partial R}{\partial t}=-qP.
\end{equation}
If an absorption line is homogeneously broadened and exact resonance
condition holds, then the SIT equations reduce to the well-known \emph{%
Sine-Gordon equation}

\begin{equation}
\frac{\partial ^{2}u}{\partial \tau \partial \zeta }+\sin u=0,  \label{eq7}
\end{equation}
where $\tau =(t-z/c)$, $\zeta =\alpha ^{\prime }z$, and $q=\partial
u/\partial \tau $.

These equations (i.e., (\ref{eq2-1} - \ref{eq2-2}), (\ref{eq5-1} - \ref
{eq5-2}), and (\ref{eq7-1} - \ref{eq7-2})) may be represented as the
condition of the integrability of some linear equations that provides to the
solution of these equations be the IST method. If we assume, that before the
arrival of an ultra-short pulse all two-level atoms are in the ground state
and after the passing of the USP all atoms recover in the initial states,
then boundary conditions are

\[
\stackunder{|\tau |\rightarrow \infty }{\lim }R(\tau ,\zeta ;\omega
_{a})=-1,\quad \stackunder{|\tau |\rightarrow \infty }{\lim }r_{1,2}(\tau
,\zeta ;\omega _{a})=0. 
\]
Both RMB-equations and SIT-equations with these boundary conditions can be
solved by the IST method in a regular way \cite{R8} - \cite{R15}. In general
case one obtain the $N$-soliton solution of these equations. This solution
represents $L_{1}$ single solitons and $L_{2}$ $\emph{breathers}$ (so that $%
N=L_{1}+2L_{2}$ ). Breather (or bion -- soliton-antisoliton bounded state)
is an extremely stable solitary wave with internal oscillations. It has the
same collision stability as ordinary solitons in both the bion-bion and the
bion-soliton collisions. It is worth to note that the Sine-Gordon equation
has the same soliton and breather solution.

It should be note that one-soliton solution of the RMB equations corresponds
the USP without carrier wave and represents a unipolar spike of
electromagnetic radiation. Sometimes these pulses are named the \emph{video} 
\emph{pulses}. Two-soliton solution of the RMB-equations describes the
collision of two video pulses in the same fashion as it was done for two
soliton solution of SIT equations by McCall-Hahn. However, the breather
solution of the RMB-equations can be used to obtain the generation of
McCall-Hahn $2\pi $-pulses. In \cite{R6} it was shown that the RMB breather
is a real solution in the form of localised pulse with internal
oscillations. Hence, this is an exact analogue of the McCall-Hahn $0\pi $%
-pulse. If the frequency of internal oscillations increases, then the
envelope of RMB breather can be described by the soliton solution of the SIT
equations to a high accuracy. Thus admittedly, the $2\pi $-pulse of
McCall-Hahn is the limiting case of the $0\pi $-pulse of the RMB equations.

\subsection{Optical solitons in fibres}

It is well known that the rate of the information transfer by means of fibre
optical communication systems (FOCS) with the mode pulse-code modulation is
limited mainly by the effect of dispersion of group velocities. The
influence of this effect can be suppressed efficiently and ideally can be
completely excluded if one uses sufficiently powerful pulses of light.
Because of the non-linear effect of self-influence such pulses in the
certain conditions are transformed in solitons and their propagation in FOCS
does not accompany by dispersive spreading.

It is important to emphasise that soliton does not exist in real
communication systems in the true sense of the word. The influence of group
velocities dispersion of the high orders, optical loss and some other
effects break a dynamic balance between the non-linear compression of pulse
and its dispersion broadening. As a result of these effects the optical
pulse suffers an envelope distortion and damping. But the distance, passed
by the soliton in a fibre considerably exceeds one, which weak pulse should
be passed in the linear regime of propagation. The experiments made with
powerful optical pulses confirm this. We can consider the soliton as a good
approach for the real non-linear pulses in a fibre under certain conditions.

The suggestion to use optical solitons for the information transfer along
the fibre was made in the works \cite{R18, R19} and it was demonstrated \cite
{R20, R21, R22}. Later on the equation describing the optical pulse
propagation in a fibre with account of only the second-order
group-velocities dispersion was received in works \cite{R23, R24} and was
investigated in \cite{R25, R26}. In the soliton theory it was known as the 
\emph{non-linear} \emph{Schr\"{o}dinger} (NLS) equation.

Let $q$ is the normalised slowly varying complex envelope of the optical
pulse defined by the following expression 
\[
E(x,y,z,t)=A_{0}q(z,t)\Psi (x,y)\exp [i(\beta _{0}z-\omega _{0}t)], 
\]
where $\beta _{0}$ is the propagation constant depending on the frequency of
the carrier wave $\omega _{0}$, $\Psi (x,y)$ is a mode function that
determines the transverse distribution of the electric field over the fibre
cross-section. Slowly varying envelope of optical pulse is governed by the
follow equation \cite{R23, R24} (NLS equation):

\begin{equation}
i\frac{\partial q}{\partial \zeta }+s\frac{\partial ^{2}q}{\partial \tau ^{2}%
}+\mu |q|^{2}q=0.  \label{eq8}
\end{equation}
Here $\zeta =z/L_{D}$, $\tau =(t-z/v_{g})t_{p0}^{-1}$ are normalised
independent variables of co-ordinate and time, accordingly, $t_{p0}$\ is a
pulse duration at $z=0$ and $v_{g}$ is the group velocity of an optical
pulse. The term in (\ref{eq8}) with the second derivative with respect to $%
\tau $ describes the pulse dispersion broadening ($s=-1$ for the normal
dispersion and $s=+1$ for the anomalous dispersion). $L_{D}$ is the
dispersion length $L_{D}=4\beta _{0}t_{p0}^{2}(|\partial ^{2}\beta /\partial
\omega ^{2}|)^{-1}$. The effective refractive index $n_{eff}$\ is defined as 
$\beta (\omega )=(\omega /c)n_{eff}$. The third term in (\ref{eq8}) is
responsible for the self-action effect. Coefficient $\mu $ is equal to the
ratio of the dispersion $L_{D}$ length to the Kerr length $L_{K}$, where $%
L_{K}=c^{2}\beta _{0}(2\pi \omega _{0}^{2}A_{0}^{2}|\chi _{R,eff}|)^{-1}$.
Here $\chi _{R,eff}$ is the effective non-linear susceptibility responsible
for the Kerr effect.

The complete integrability of the NLS was established in the classical works 
\cite{R25, R26, R27}. Besides the multi-soliton solutions the NLS equations
has the new type of the solutions named multiple-pole solutions \cite{R8}.
These solutions correspond to the multiple points of a discrete spectrum of
the Zakharov-Shabat problem in IST method. It is difficult to realise these
solutions in practice because any small stir of the initial conditions will
remove degeneration in the multiple points of a discrete spectrum. It means
that only the exclusive initial optical pulses can be transformed into
multiple-pole optical solitons. Different properties of the NLS equations
are described in the excellent books \cite{R29} - \cite{R33}.

\subsection{Interaction of three waves}

One of the broadest classes of phenomena in non-linear optics is the
transformation of the frequency of an electromagnetic radiation propagating
in the non-linear medium. Harmonics generation of the fundamental wave
(pump), sum-frequency and difference-frequency mixing are classified among
these phenomena \cite{R34}. Under sufficiently high intensity of a pump the
polarisation of a medium is not a linear function of the electric field
strength of the wave. If the frequencies of an electromagnetic field are not
in resonance with atomic transition frequencies, one can use a standard
perturbation theory to reveal this dependency. So we can expand polarisation 
$\vec{P}$ in a power series of electrical field strength. The coefficients
of this series are tensors of the $n$th rank $\hat{\chi}^{(n)}$, named as
non-linear susceptibility, describe different processes of the
electromagnetic waves interaction. The non-linear effects, described by the $%
n$th rank tensors of non-linear susceptibility, are often interpreted as the
interaction of the $(n+1)$ waves.

Let the non-linear characteristics of a medium be described by non-linear
susceptibility of the second order $\hat{\chi}^{(2)}$. They call it
quadratic non-linear media. Let the waves with the carrier frequencies $%
\omega _{1}$ and $\omega _{2}$ propagate along $z$ axis. As the polarisation
is the non-linear (quadratic) function of the electrical field strengths,
the waves with carrier frequencies $\omega =\omega _{1}\pm \omega _{2}$, $%
\omega =2\omega _{1}$ and $\omega =2\omega _{2}$\ appear in such medium.
These waves, in their turn, can cause generation of new waves with the
frequencies $\omega =2\omega _{1}\pm \omega _{2}$, $\omega =\omega _{1}\pm
2\omega _{2}$, and so on. But in a dispersive medium all these processes are
not equally efficient. There is a condition of phase matching, which selects
a certain type of interaction of three waves, leaving all other unaffected.
Sometimes such phase matching takes place for the waves propagating in the
same direction. In this case they say about collinear parametric
interaction. In this case the distance where the interaction of waves occurs
can be made sufficiently long and, consequently, the effective frequency
transformation will take place. On the contrary, when the phase matching is
achievable only for the waves propagating in different directions, their
interaction occurs only in the field of overlapping of the wave beams.
Non-collinear parametric interaction is worth to draw special attention as
the number of interesting results concerning the integrability of three-wave
mixing equations have been found.

Let $\mathcal{E}_{1},\mathcal{E}_{2}$\ and $\mathcal{E}_{3}$\ are the slowly
varying envelopes of the interacting pulses. Let us consider the situation
when only the collinear propagating wave with sum-frequency or difference
frequency is generated. In the slowly varying envelopes and phases
approximation, the system of equations describing the interaction of the
three waves can be written in the following in unified form \cite{R35, R36}:

\[
\left( \frac{\partial }{\partial z}+\frac{1}{v_{1}}\frac{\partial }{\partial
t}\right) q_{1}=i\sigma q_{2}^{\ast }q_{3}^{\ast }\exp (+i\Delta kz) 
\]

\begin{equation}
\left( \frac{\partial }{\partial z}+\frac{1}{v_{2}}\frac{\partial }{\partial
t}\right) q_{2}=i\sigma q_{3}^{\ast }q_{1}^{\ast }\exp (+i\Delta kz)
\label{eq9}
\end{equation}

\[
\left( \frac{\partial }{\partial z}+\frac{1}{v_{3}}\frac{\partial }{\partial
t}\right) q_{3}=-i\sigma q_{1}^{\ast }q_{2}^{\ast }\exp (+i\Delta kz) 
\]
where $\sigma =\sqrt{\gamma _{1}\gamma _{2}\gamma _{3}}$, and $\gamma
_{n}=4\pi \omega _{n}\chi ^{(2)}(\omega _{1},\omega _{2})/cn(\omega _{n})$, $%
n=1,2,3$. In these equations we used $\Delta k=k_{3}-(k_{1}+k_{2})$, for a
sum-frequency mixing process $\omega =\omega _{1}+\omega _{2}$, and $\Delta
k=k_{3}-(k_{1}-k_{2})$, for difference-frequency mixing $\omega =\omega
_{1}-\omega _{2}$. Here $v_{n}$\ are the group velocities of the
corresponding wave. The effects of group-velocity dispersion are neglected.
The uniformization variables $q_{1,2,3}$\ related with\ $\mathcal{E}_{1},%
\mathcal{E}_{2}$\ and $\mathcal{E}_{3}$ as follows $\mathcal{E}_{1}=\sqrt{%
\gamma _{1}}q_{1}$, $\mathcal{E}_{2}=\sqrt{\gamma _{2}}q_{2}$\ and $\mathcal{%
E}_{3}=\sqrt{\gamma _{3}}q_{3}^{\ast }$, , for the case of sum-frequency
mixing, and $\mathcal{E}_{1}=-\sqrt{\gamma _{1}}q_{3}$, $\mathcal{E}_{1}=%
\sqrt{\gamma _{2}}q_{2}^{\ast }$, $\mathcal{E}_{3}=\sqrt{\gamma _{3}}%
q_{1}^{\ast }$\ for the case of difference-frequency mixing (the indices at
the velocities of the first and the third waves should be changed in this
case).

We would like to point out that the resonance Raman scattering under
condition of weak variation of the energy levels population of a medium and
the scattering of optical waves by an acoustical wave can be considered from
one position as the specific realisations of the three-wave interaction. In
both cases the systems of equations, describing these processes, can be
transformed into one universal system (\ref{eq9}). Furthermore, one can
demonstrate that under certain conditions the reduced Maxwell-Bloch
equations appear here, which describe the propagation of the ultra-short
pulse in a resonance medium. Due to this property the Raman scattering can
be analysed in terms of IST method.

The equations for 3-wave interaction (\ref{eq9}) permit both the infinite
number of conservation laws and the B\"{a}cklund transformation. They can be
presented as the Hamiltonian equations by employing the $r$-matrix. It was
shown that these equations pass the Painleve test and that there is a class
of self-similar solutions, expressed in terms of Painleve transcendents P-V
and P-VI.

It is remarkable that the system of equations, describing the parametric
interaction of three waves, can be solved by IST method for a 3D case \cite
{R37, R38}, whereas the most of soliton equations are one-dimensional ones.

It should be emphasised that the 3-wave interaction gives an example of the
dispersionless propagation of the non-linear waves. They often say that
soliton is the result of the compensation of dispersion broadening and
non-linear compression of the wave packet. The 3-wave interaction just
demonstrates the narrowness of this statement. Due to the absence of the
phase and group velocities dispersion the solitons in this process do not
detach from the non-soliton part of the solution (which is often named
radiation). It seems very difficult to study the process analytically, so
one has to confine the investigation to some particular solutions. The
non-collinear second harmonic generation provides an example of a specific
case of the 3-wave interaction, where an exact solution was found without
employing the IST method. The solution obtained explicitly illustrates the
non-separability of the soliton and non-soliton parts of the solution of
these three wave interaction equations.

\section{New examples of the integrable systems}

\subsection{Generalisation of the self-induced transparency theory}

The development of the SIT theory is characterised by its going beyond the
framework of two-level approximation and by the spectral composition of the
USP field becoming more complex. The latter means that the resonant medium
interacts with radiation containing several carrier frequencies. Besides,
such factors as the direct interaction between resonant atoms, non-linear
properties of the dielectric doped with resonant atoms and polarisation of
the electromagnetic field should be taken into account.

Let the optical pulse propagate along axis $z$ and the electric field
strength be presented in the form\ $\vec{E}=\mathcal{\vec{E}}(t,z)\exp
(-i\omega _{0}t+ik_{0}z)+c.c.$\ The carrier frequency is in resonance with
the frequency $\omega _{21}=(W_{2}-W_{1})/\hbar $ of an atomic transition $%
j_{2}\rightarrow j_{1}$ between energy levels $W_{2}$ and $W_{1}$
degenerated over the projections $m$ and $l$ of the total angular momenta $%
j_{1}$ and $j_{2}$. In a general case the evolution of the envelope of the
USP and states of the resonant medium are described by the system of
equations for which the exact solution is not known in the case of arbitrary
values$j_{1}$ and $j_{2}$. But the certain choice of transitions $%
j_{1}=0\leftrightarrow j_{2}=1$, $j_{1}=1\leftrightarrow j_{2}=1$, and $%
j_{1}=1/2\leftrightarrow j_{2}=1/2$ makes this system of equations exactly
integrable. Its solution can be obtained by the inverse scattering transform
method, as it was shown in \cite{R39, R40, R41}. \emph{Generalised reduced
self-induced transparency equations} (GSIT equations), for transitions $%
j_{1}=0\leftrightarrow j_{2}=1$, $j_{1}=1\leftrightarrow j_{2}=1$, can be
represented in the homogeneous form

\[
\frac{\partial q_{j}}{\partial \zeta }=-i\stackunder{a=1,2}{\sum }\beta
_{a}\left\langle P_{j}^{(a)}\right\rangle , 
\]

\begin{equation}
i\left( \frac{\partial }{\partial \tau }-i\Delta \omega t_{p0}\right)
P_{j}^{(a)}=\stackunder{l}{\sum }q_{l}M_{lj}^{(a)}-q_{j}N^{(a)},
\label{eq10}
\end{equation}

\[
\frac{\partial }{\partial \tau }M_{jl}^{(a)}=-i\left( q_{l}^{\ast
}P_{j}^{(a)}-q_{l}P_{j}^{(a)\ast }\right) ,\quad \frac{\partial }{\partial
\tau }N^{(a)}=i\stackunder{j}{\sum }\left( q_{j}^{\ast
}P_{j}^{(a)}-q_{j}P_{j}^{(a)\ast }\right) . 
\]

If we consider transition $j_{1}=0\rightarrow j_{2}=1$, then in (\ref{eq10})
we assign $q_{j}=d\mathcal{E}^{j}t_{p0}/\hbar $, $\beta _{1}=1$ and $\beta
_{2}=0$. For transition $j_{1}=1\rightarrow j_{2}=0$\ we have $q_{j}=d%
\mathcal{E}^{j}t_{p0}/\hbar $, $\beta _{1}=0$ and $\beta _{2}=1$ . Finally,
for transition $j_{1}=1\rightarrow j_{2}=1$\ we have $q_{j}=jd\mathcal{E}%
^{j}t_{p0}/\hbar \sqrt{2}$, $\beta _{1}=\beta _{2}=1/2$. Everywhere here the
sub-index and upper index take the values $j=\pm 1$. The slowly varying
envelopes of matrix elements of the density matrix $\hat{\rho}$\ are
determined as $P_{j}^{(1)}=\left\langle j_{2},0|\hat{\rho}%
|j_{1},j\right\rangle $, $P_{j}^{(2)}=\left\langle j_{2},-j|\hat{\rho}%
|j_{1},0\right\rangle $, $N^{(1)}=\left\langle j_{2},0|\hat{\rho}%
|j_{2},0\right\rangle $, $N^{(2)}=-\left\langle j_{1},0|\hat{\rho}%
|j_{1},0\right\rangle $, $M_{jl}^{(1)}=\left\langle j_{2},j|\hat{\rho}%
|j_{1},l\right\rangle $, $M_{jl}^{(2)}=\left\langle j_{2},-l|\hat{\rho}%
|j_{1},-j\right\rangle $. As we are well aware, GSIT equations are the
zero-curvature condition in the vector expansion IST method for the AKNS
hierarchy. The spectral problem with such a sort of IST was first reported
by Manakov \cite{R42} to describe the self- focusing of the polarised light
beams. The expressions for $N$-soliton solutions, breathers and B\"{a}cklund
transformation have been found in \cite{R43}.

There are versions of the GSIT equations describing USP propagation in
three-level medium. In the simplest case of this model the resonance levels
have $V$ and $\Lambda $ configurations. It was determined \cite{R44, R45}
that, if the oscillator forces for every transition in a $V$ or $\Lambda $
configuration are equal, then a two-frequency pulse (characterised by two
different frequencies of the carrier wave) is able to propagate in such a
medium without the envelope distortion. An ultra-short pulse of the kind was
called \emph{simulton} \cite{R46}. The simultons are single-soliton
solutions of the above GSIT equations. At the same time solutions occur
which are responsible for the propagation and collisions of simultons. The
oscillating simultons (colour breathers) are the two-frequency
generalisations of $0\pi $-pulse by McCall-Hahn. It is worth noting at this
point that a simulton is generally unstable with respect to transformation
into one-frequency $2\pi $-pulse and may remain as two-frequency pulse only
for a special choice of resonance level populations.

It should be mentioned that consideration of the polarised USP propagation
ion three-level medium leads to \emph{matrix variants of the GSIT-equations}%
. For more details, see Ref. \cite{R47, R48}. The matrix expansion IST
method for the AKNS hierarchy (or matrix Manakov spectral problem) there
occurs.

\subsection{Femtosecond optical solitons in fibres}

In order to describe the non-linear phenomena associated with femtosecond
optical solitons in non-linear fibres, \emph{the higher-order non-linear
Schr\"{o}dinger} (HNLS) equation has been proposed \cite{R49, R50} (see also 
\cite{R51}). Consider the following generalisation of the equation (\ref{eq8}%
)

\begin{equation}
i\frac{\partial q}{\partial \zeta }+s\frac{\partial ^{2}q}{\partial \tau ^{2}%
}+\mu |q|^{2}q+i\left( \eta _{3}\frac{\partial ^{3}q}{\partial \tau ^{3}}%
+\mu _{2}|q|^{2}\frac{\partial q}{\partial \tau }+\mu _{3}q\frac{\partial
|q|^{2}}{\partial \tau }\right) =0.  \label{eq11}
\end{equation}
The parameter $\eta _{3}$ corresponds to the third-order group-velocity
dispersion, parameters $\mu _{2}$\ and $\mu _{3}$ represent the two inertial
contributions to the non-linear polarisation, i.e., Raman self- scattering
and self-steeping formation. If\ $\eta _{3}=0$, $\mu _{2}$=$\mu _{3}$=1, and 
$\mu $=0, then the HNLS (\ref{eq11}) reduced into \emph{derivative
non-linear Schr\"{o}dinger }(DNLS) equation

\begin{equation}
i\frac{\partial q}{\partial \zeta }+\frac{\partial }{\partial \tau }\left( s%
\frac{\partial q}{\partial \tau }+i|q|^{2}q\right) =0.  \label{eq12}
\end{equation}
This equation is completely integrable one \cite{R52}. There are soliton and
multi-soliton solutions, which can be obtained in the framework of IST
method. Notice that the condition $\mu $=0 is not essential for reduction of
the HNLS into the integrable equation, i.e., resulting modified DNLS
equation is completely integrable too.

If $\eta _{3}$= 1, $\mu _{2}=\pm 6$, $\mu _{3}$= 0, then the equation (\ref
{eq11}) is reduced to \emph{the Hirota equations} \cite{R53} , \bigskip 
\begin{equation}
i\frac{\partial q}{\partial \zeta }+s\frac{\partial ^{2}q}{\partial \tau ^{2}%
}+\mu |q|^{2}q+i\frac{\partial ^{3}q}{\partial \tau ^{3}}\pm 6i|q|^{2}\frac{%
\partial q}{\partial \tau }=0,  \label{eq13}
\end{equation}
which represents other example of completely integrable equation.

In general, equation (\ref{eq11}) may not be completely integrable. However,
if we suppose that $\eta _{3}=\varepsilon $, $\mu _{2}=6\varepsilon $, $\mu
_{3}=3\varepsilon $, and $\mu =2s$, then the HNLS equation can be reduced to
the other one -- \emph{Sasa-Satsuma equation}. In this case in terms of new
variables $\xi =\tau -s^{2}\zeta /3\varepsilon $, $q(\tau ,\zeta )=u(\xi
,\zeta )\exp (is\tau /3\varepsilon +2is^{3}\zeta /27\varepsilon ^{2})$\ the
equation (\ref{eq11}) takes the form

\begin{equation}
\frac{\partial u}{\partial \zeta }+\varepsilon \left( \frac{\partial ^{3}u}{%
\partial \xi ^{3}}+6|u|^{2}\frac{\partial u}{\partial \xi }+3u\frac{\partial
|u|^{2}}{\partial \xi }\right) =0.  \label{eq14}
\end{equation}
The equation (\ref{eq14}) has been considered in (\cite{R54}), where it was
shown that this equation could be solvable by means of IST method.

\subsection{Vector optical solitons}

In the general case the \emph{vector soliton} is the soliton solution of the
non-linear system of the evolution equations, which one can be presented as
the one-dimension array. For example, the vector soliton of the NLS equation
is the solution of the following vector equation

\begin{equation}
i\frac{\partial \vec{q}}{\partial \zeta }+\frac{1}{2}\frac{\partial ^{2}\vec{%
q}}{\partial \tau ^{2}}+(\vec{q}\vec{q}^{\ast })\vec{q}=0,  \label{eq15}
\end{equation}
where $\vec{q}=\{q_{1},q_{2},...,q_{M}\}$. Hereafter, this equation will
referred to as the \emph{v-NLS equation}. The vector index of this soliton
can be resulted from the different physical origin.

We have only one example where the v-NLS equation is the completely
integrable one. This equation is embedded into AKNS hierarchy that allows to
exploit the IST method to find the soliton solution. The suitable spectral
problem was found by Manakov \cite{R42}.

As the simplest vector generation of the DNLS equation one can write the
standard form of the \emph{v-DNLS equation}

\begin{equation}
i\frac{\partial \vec{q}}{\partial \zeta }+\frac{\partial }{\partial \tau }%
\left( \frac{\partial \vec{q}}{\partial \tau }-i\varepsilon (\vec{q}\vec{q}%
^{\ast })\vec{q}\right) =0.  \label{eq16}
\end{equation}

Other example of the vector non-linear waves arises under consideration
ultra-short optical pulse propagation in birefringent fibres with higher
order effects like the third order dispersion of group velocities, Kerr
dispersion, and stimulated Raman scattering. There is \emph{two-component
generalisation of the Sasa-Satsuma equation (}\ref{eq14}) \cite{R55}:

\begin{equation}
\frac{\partial \vec{q}}{\partial \zeta }+\varepsilon \left( \frac{\partial
^{3}\vec{q}}{\partial \xi ^{3}}+6(\vec{q}\vec{q}^{\ast })\frac{\partial \vec{%
q}}{\partial \xi }+3\vec{q}\frac{\partial (\vec{q}\vec{q}^{\ast })}{\partial
\xi }\right) =0.  \label{eq17}
\end{equation}
In \cite{R55} the three-component generalisation of this equation was
considered too. It was shown that these equations have the zero-curvature
representation and can be solved by IST method. However, the exact soliton
solutions more simply to obtain using the Darboux- B\"{a}cklund
transformation.

\subsection{Extremely short pulse propagation in non-resonant medium}

The recent progress in the field of generation of femtosecond pulses has
made it necessary to revise theoretical models of their propagation in a
non-linear dispersive medium. It is interesting to find a method to describe
an USP evolution without the use of slowly varying envelope approximation.
The simplest way to do this is to combine the wave equation for
electromagnetic field with the equations specifying the changes in the state
of the medium.

In the case of resonant medium we have the resonance transition frequency as
a scale parameter for time. When pulse duration $t_{p}$\ obeys the
inequality $t_{p}\omega _{a}\gg 1$, the slowly varying envelope
approximation is adequate to describe the pulse propagation. On the
contrary, if $t_{p}\omega _{a}\ll 1$\ we can use at least the unidirectional
propagation approximation. The ratio $\varepsilon =\omega _{R}/\omega _{a}$,
where $\omega _{R}$ is the Rabi frequency (i.e., $\omega _{R}=d\max
|E|/\hbar $), provides a new parameter. Let the USP amplitude (i.e., $E$) be
of such magnitude that the Rabi frequency $\omega _{R}$ is small compared
with the minimum of atomic transition frequency. That means $\varepsilon
=\omega _{R}/\omega _{a}$\ is a small parameter. Then we can attempt to
solve the Bloch equations (\ref{eq2-2}) approximately and thus obtain an
approximate equations of the USP electric field strength without the
assumption of slowly varying envelope \cite{R56}. The Maxwell equations
under unidirectional wave approximation can be reduced into

\begin{equation}
\frac{\partial E}{\partial z}+\frac{1}{c}\frac{\partial E}{\partial t}=-%
\frac{2\pi n_{A}d}{c}\left\langle \frac{\partial P_{1}}{\partial t}%
\right\rangle ,  \label{eq18}
\end{equation}
where polarisation of the ensemble of two-level atoms at the third order of $%
\varepsilon $ is

\begin{equation}
\left\langle P_{1}\right\rangle =\left\langle \frac{2d}{\hbar \omega _{a}}%
\right\rangle E-\left\langle \frac{2d}{\hbar \omega _{a}^{3}}\right\rangle 
\frac{\partial ^{2}E}{\partial t^{2}}-\left\langle \frac{4|d|^{2}d}{\hbar
^{3}\omega _{a}^{3}}\right\rangle E^{3}.  \label{eq19}
\end{equation}
Substitution of the expression for polarisation into equation (\ref{eq18})
in terms of new variables $\tau =|b|z$, $\zeta =t-z/V$, $u(\tau ,\zeta
)=-(a/6b)^{1/2}E(z,t)$ yields the \emph{modified Korteweg-de Vries equation}
(mKdV)

\begin{equation}
\frac{\partial u}{\partial \tau }+6u^{2}\frac{\partial u}{\partial \zeta }+%
\frac{\partial ^{3}u}{\partial \zeta ^{3}}=0.  \label{eq20}
\end{equation}
Here parameters $a=\left\langle 24\pi n_{A}|d|^{4}/c\hbar ^{3}\omega
_{a}\right\rangle $, $b=\left\langle 4\pi n_{A}|d|^{2}/c\hbar ^{3}\omega
_{a}\right\rangle $ were introduced. The expression $V^{-1}=c^{-1}[1+\left%
\langle 4\pi n_{A}|d|^{2}/\hbar \omega _{a}\right\rangle ]$ defines the
re-normalised velocity of the USP propagation. As it is known \cite{R57}
this equation is a completely integrable, and its solutions can be found by
the IST method \cite{R31, R32, R33}. There are another non-linear equations
describing extremely short pulse propagation \cite{R58}, however, they have
not soliton solutions.

\subsection{SIT in a Kerr-type non-linear medium}

There are intensive investigations of the non-linear pulse propagation in
the optical fibres \cite{R29, R30}. However, practically all materials used
for fibre fabrication contain impurities that contribute to the absorption
spectrum of the fibres. The losses due to the resonant absorption decrease
if the frequency of the carrier wave is located within the window of
transparency of glass fibre. Another means to decrease losses is to make
pulse duration shorter than the characteristic relaxation times of the
resonant states or in other words to make optical pulses ultra-short. In
this case the well-known self-induced transparency phenomenon can be
expected to arise.

It is known that the non-linear Schr\"{o}dinger equation, which is used to
describe optical solitons in non-linear monomode optical fibre, is
completely integrable \cite{R31, R32, R33}. The reduced Maxwell-Bloch
equations or their generalisations considered hereafter, as RMB-equations
are completely integrable too. The IST method for both the NLS and the
RMB-equations enabling to solve some non-linear evolution equations is based
on the same spectral problem. The model of the USP propagation in a
Kerr-type non-linear medium doped by resonant impurity atoms incorporates
both of these systems. But there are no reasons for the resulting system of
equations to possess a complete integrability.

The evolution of the USP propagating in a non-linear monomode optical fibre
in $z$- direction is described by the equations, which generalise the
Maxwell-Bloch equations \cite{R59}. We could name them the non-linear
Schr\"{o}dinger and Bloch equations.

\begin{equation}
i\frac{\partial q}{\partial \zeta }+s\frac{\partial ^{2}q}{\partial \tau ^{2}%
}+\mu |q|^{2}q+a\left\langle p\right\rangle =0,  \label{eq21-1}
\end{equation}

\begin{equation}
\frac{\partial p}{\partial \tau }=i\delta p+2ifR_{3},\quad \frac{\partial
R_{3}}{\partial \tau }=if\left( q^{\ast }p-qp^{\ast }\right) ,
\label{eq21-2}
\end{equation}
where $q$ is the normalised slowly varying complex envelope of the USP
defined by the following expression

\[
E(x,y,z,t)=A_{0}q(z,t)\Psi (x,y)\exp [i(\beta _{0}z-\omega _{0}t)], 
\]
\ $\Psi (x,y)$ is a mode function that determines the transverse
distribution of the electric field over the fibre cross-section. Here $\zeta
=z/L_{D}$, $\tau =(t-z/v_{g})t_{p0}^{-1}$, are normalised independent
variables of co-ordinate and time, accordingly, $t_{p0}$\ is a pulse
duration at $z$ = 0 and $v_{g}$ is the USP propagation group velocity. The
interaction of the radiation with the resonant impurities is characterised
by the dimensionless constant $f=\bar{d}A_{0}t_{p0}/\hbar $, where $\bar{d}$
is an effective matrix element of the dipole transition between the resonant
states. The coefficient $a$ is expressed in terms of the dispersion length $%
L_{D}$\ and the resonant absorption length $L_{a}$ \cite{R48} as $%
a=L_{D}L_{a}^{-1}f^{-1}$, where $L_{a}=c\hbar n_{eff}\left( 2\pi \omega
_{0}n_{A}\bar{d}^{2}t_{p0}\right) ^{-1}$ .

Now let us consider ultra-short pulses propagation in a Kerr-type dispersive
medium when the transition between energy levels of the impurity atoms are
degenerated over the orientations of the total angular momentum $j_{1}$\ and 
$j_{2}$. The same system of equations appears when fibre contains
three-level impurity atoms, so that formally we could speak about optical
vector solitons in a general case. Let the electric field strength write as

\[
E^{(j)}(x,y,z,t)=A_{0}q_{j}(z,t)\Psi (x,y)\exp [i(\beta _{0}z-\omega
_{0}t)]. 
\]
The equations for the normalised envelope and the variables of the atomic
resonant system can be written in a unified form as it was done above

\[
i\frac{\partial q_{j}}{\partial \zeta }+s\frac{\partial ^{2}q_{j}}{\partial
\tau ^{2}}+\mu |\vec{q}|^{2}q_{j}-a\left\langle P_{j}\right\rangle =0, 
\]

\begin{equation}
\frac{\partial P_{j}^{(a)}}{\partial \tau }-i\delta P_{j}^{(a)}=-if\left( 
\stackunder{l}{\sum }q_{l}M_{lj}^{(a)}-q_{j}N^{(a)}\right) ,  \label{eq22}
\end{equation}

\[
\frac{\partial }{\partial \tau }M_{jl}^{(a)}=-if\left( q_{l}^{\ast
}P_{j}^{(a)}-q_{l}P_{j}^{(a)\ast }\right) ,\quad \frac{\partial }{\partial
\tau }N^{(a)}=if\stackunder{j}{\sum }\left( q_{j}^{\ast
}P_{j}^{(a)}-q_{j}P_{j}^{(a)\ast }\right) 
\]
where $j$ and $l$ mark the spherical components of the vectors, and $%
P_{j}=\sum_{a}\beta _{a}P_{j}^{(a)}$. The variables into the Bloch equations
in the system (\ref{eq22}) have been determined above.

It has been found \cite{R48} that both system of equations (\ref{eq21-1} - 
\ref{eq21-2})\ and (\ref{eq22}) are integrable ones only under condition $%
L_{D}L_{K}^{-1}=2f^{2}$. This condition implies that the soliton of the
self-induced transparency should simultaneously be also a soliton of the NLS
equation. To put this another way, the amplitude and duration of the $2\pi $%
-pulse should precisely be of such values that the corresponding self-action
(due to the high-frequency Kerr effect) would lead to complete compensation
of the dispersion broadening of the USP. Thereby, the existence of the
optical solitons in a fibre doped with resonant impurities is restricted.

Yet another example of integrable model describing USP propagation in fibre
was proposed in \cite{R60}. There the NLS equation (\ref{eq21-1}) has been
replaced by the Hirota-like equation:

\begin{equation}
i\frac{\partial q}{\partial \zeta }+\frac{1}{2}\frac{\partial ^{2}q}{%
\partial \tau ^{2}}+|q|^{2}q+i\varepsilon \left( \frac{\partial ^{3}q}{%
\partial \tau ^{3}}+3|q|^{2}\frac{\partial q}{\partial \tau }+\frac{3}{2}q%
\frac{\partial |q|^{2}}{\partial \tau }\right) +a\left\langle p\right\rangle
=0.  \label{eq23}
\end{equation}
The resulting system of equation admits the Painleve property for certain
relation between the physical parameters involved in the model. Also the
zero-curvature representation of this system was explicitly found.

\section{Conclusion}

In conclusion it will be considered some problems related with an integrable
models and developments of the ideas mentioned above.

In the traditional scheme of the IST method the spectral parameter of the
auxiliary linear problem is considered as a constant. The generation of the
IST method, where the spectral parameter depends on time and space
co-ordinate was proposed in \cite{R61}. Non-linear equations arising in this
approach include the explicit dependence on co-ordinates. In framework of
this method the generation of the Maxwell-Bloch equations was constructed.
This new system describes evolution of the USP in two-level medium wherein
the pumping of the excited states of atoms operated continuously.

Propagation of the USP in a long two-level amplifier was considered in \cite
{R62, R63}. The IST method was applied to the SIT equations in order to
obtain its uniform asymptotic solution at long distance. It was shown that
the amplified pulse is always of a quasi self-similar nature. In the
neighbourhood of the wave front USP is described by the Painleve equation,
whereas far from the front the solution goes into the rapidly oscillating
self-similar regime.

The SIT equations has been employed \cite{R64} to describe the phenomenon of
superfluorescence (i.e., short pulse generation from initial fluctuations of
polarisation in a high inverted of the population in two-level medium).
Contrary to the self-induced transparency, the superfluorescence pulse is
generating from an unstable state and IST method is to be re-formulated for
boundary problem at the half $t$-axis. It was done in \cite{R64}.

Progress in laser physics has made possible the generation of intense
ultra-short pulses to produce the multi-photon processes. The simplest one
develops when the resonant matter interacts with a pair of USPs of different
carrier frequencies $\omega _{1}$\ and $\omega _{2}$\ so that $\omega
_{1}\pm $\ $\omega _{2}$ coincides with a transition frequency $\omega _{21}$%
. The non-linear processes are two-photon absorption and Raman scattering,
respectively. The coherent pulse propagation takes place under both
resonance conditions when the pulse duration is much less then relaxation
times. This process is similar to SIT and it was referred to as \emph{%
two-photon self-induced transparency} (TPSIT). This phenomenon can be
described in framework of the generalised Maxwell-Bloch equations \cite{R65,
R66}. These equations were converted into new system of equation for new
variables, which are quadratic functions of initial slowly varying envelopes
of the electromagnetic fields \cite{R66, R67}. As was shown by Kaup and
Steudel \cite{R67, R68} under certain conditions these equations can be
solved by using modified IST method.

It is interesting that both McCall-Hahn equations in the sharp-line limit
and exact resonance and GSIT equations can be mapped into the model of \emph{%
principal chiral fields}. The Kaup-Steudel equations are related with the
same model \cite{R69}. Recently the authors of \cite{R70} have investigated
the GSIT equations and found the hidden non-Abelian group structure of these
equations in the case of multi-level resonant medium. They have discovered
that a non-degenerate two-level system of self-induced transparency is
associated with symmetric space $G/H=SU(2)/U(1)$\ while three-level $V$- or $%
\Lambda $-systems are associated with $G/H=SU(3)/U(2)$. The same symmetric
space is associated with the degenerate two-level system of SIT in the case
of transitions $j_{1}=1\rightarrow j_{2}=0$\ and $j_{1}=0\rightarrow j_{2}=1$%
. When one considers the transition $j_{1}=1/2\rightarrow j_{2}=1/2$, the
GSIT equations are associated with $G/H=(SU(3)/U(2))^{2}$. There are many
complex aspects related to the degeneration of the energy levels in a
three-level system of SIT. For instance, the transitions between states $%
j_{a}=j_{c}=0,j_{b}=1$\ (or $j_{a}=j_{c}=1,j_{b}=0$ ) are associated with
the symmetric space $G/H=SU(4)/S(U(2)\times U(2))$ (accordingly $%
G/H=SU(5)/U(4)$ ).

The degenerate four-wave interaction consisting of two counter-propagating
pulses in a cubic non-linear medium, with arbitrary polarisations which can
vary through each pulse has been considered in \cite{R71, R72}. The group
velocity dispersion was neglected that results in the intensity envelopes
propagating unchanged from their initial forms. However, the polarisation
evolves according to the non-linear interaction and it is found it can
exhibit soliton behaviour. The problem is best approached if one introduce
the following vectors (i.e., Stokes vectors) 
\[
\mathbf{S}^{(\pm )}=\left\{ q_{x}^{(\pm )\ast }q_{y}^{(\pm )}+c.c.,\quad
iq_{x}^{(\pm )\ast }q_{y}^{(\pm )}+c.c.,\quad |q_{x}^{(\pm
)}|^{2}-|q_{y}^{(\pm )}|^{2}\right\} 
\]
where $q_{x,y}^{(\pm )}$ are the transverse normalised slowly varying
envelopes of the pulses propagating in the $+\mathbf{\hat{z}}$ and $-\mathbf{%
\hat{z}}$ directions, respectively. By using the cone co-ordinate $2x_{\pm
}=\mp (z\pm ct)$, the evolution equations for can be re-written as the
equation of motion for chiral field on the group $O(3):$

\begin{equation}
\partial _{+}\mathbf{S}^{(+)}=\mathbf{S}^{(+)}\times \hat{J}\mathbf{S}%
^{(-)},\quad \partial _{-}\mathbf{S}^{(-)}=-\mathbf{S}^{(-)}\times \hat{J}%
\mathbf{S}^{(+)}  \label{as}
\end{equation}
where $\hat{J}$ diagonal matrix is defined by constants of the model. These
equations are integrable by IST method \cite{R73, R74}. It should be pointed
that in the anisotropic case (i.e., matrix elements of are different) the
equations (\ref{as}) describe the \emph{domain wall} (DW) in the
configuration of the Stokes vectors. As far as we know it is the first
example of domain structure in non-linear optics, where DW's represent space
regions with different stable polarisation states of the interacting optical
waves. The Hamiltonian achieves the minimum on these optical domains, and
DW's are the regions of polarisation switching. Recently, study of the DW's
was considered in systems of coupled NLS equations governing propagation of
light in non-linear fibres \cite{R75}. An important point is that
group-velocity dispersion was taken into account. The two types of the DW
were found: the one between different elliptic polarisations in the bimodal
fibre and a dark soliton in one core of the dual-core coupler. However, both
the model considered in \cite{R75} and models quoted there are not
integrable ones.

\section*{Acknowledgment}

I am grateful to Prof. S.O. Elyutin and Dr. A.M. Basharov for enlightening
discussions. Also financial support from INTAS (grant No: 96-0339) is
acknowledged.

\end{document}